# Sub-Poissonian shot noise in CoFeB/MgO/CoFeB-based magnetic tunneling junctions

Tomonori Arakawa[1], Koji Sekiguchi[1], Shuji Nakamura[1], Kensaku Chida[1], Yoshitaka Nishihara[1], Daichi Chiba[1], Kensuke Kobayashi[1 a)], Akio Fukushima[2], Shinji Yuasa[2] and Teruo Ono[1]

[1] *Institute for Chemical Research, Kyoto-University, Uji, Kyoto 611-0011, Japan*
[2] *Spintronics Research Center, Advanced Industrial Science and Technology (AIST), AIST Tsukuba Central 2, Tsukuba, Ibaraki 305-8568, Japan.*

We measured the shot noise in the CoFeB/MgO/CoFeB-based magnetic tunneling junctions with a high tunneling magnetoresistance ratio (over 200 % at 3 K). Although the Fano factor in the anti-parallel configuration is close to unity, it is observed to be typically 0.91±0.01 in the parallel configuration. It indicates the sub-Poissonian process of the electron tunneling in the parallel configuration due to the relevance of the spin-dependent coherent transport in the low bias regime.

a) Electronic mail: kensuke@scl.kyoto-u.ac.jp

The tunneling magnetoresistance (TMR) effect in magnetic tunneling junctions (MTJs), which consist of a tunnel barrier sandwiched by two ferromagnetic electrodes, is a typical example of the spin-dependent electron transport.[1] When the tunnel barrier is composed of the amorphous $Al_2O_3$,[2] the electron tunneling can be explained by the conventional Julliere's model,[3] whereas the coherent tunneling is theoretically discussed to play a central role in MTJs with the crystalline MgO barriers.[4-7] To investigate the spin-dependent transport in MTJs, most studies so far have focused only on their *I-V* characteristics. However, the shot noise measurement is expected to provide further insight into the mechanism of the electron transport.[8]

Generally, when the current $I$ is fed to a tunnel junction, the current noise $S_I$ due to the shot noise occurs. $S_I$, which is frequency-independent, can be expressed as $S_I = 2eIF$ (in the zero-temperature limit) with Fano factor $F$. In normal-insulator-normal junctions, $F=1$ is established, which means that the electron tunneling events through the barrier are independent of each other (the Poissonian process).[9] Although several papers reported the 1/f-noise properties in MTJs,[10-14] very few studies have been reported on the Fano factor. One of them reported $F$~1 in $Al_2O_3$-based MTJs,[15] while they also reported the sub-Poissonian shot noise, namely a Fano factor less than 1.[16] A similar observation was reported in $Al_2O_3$-based MTJs.[17] Regarding MgO-based MTJs, we reported that the Fano factor is close to unity[18] in agreement with the pioneering work by Guerrero *et al.*[19]

In this study, we report the shot noise in the well-crystalline MgO-based MTJs with the experimental accuracy below 1 %, while the experimental accuracy of the previous work[18] was limited within ~5 %. The MgO barrier is much thinner than those in our previous study to address the relevance of the coherent transport. We observed reduced Fano factor (typically, $F=0.91 \pm 0.01$) in the parallel (P) configuration, while $F$ in the anti-parallel (AP) configuration is close to unity. The reduction is observed to be independent of the sample temperature and the magnetic field.

The present MTJs are the multilayer stacks of the buffer / PtMn(15) /CoFe(2.5) / Ru(0.85) / CoFeB(3) / MgO(1.05) / CoFeB(2) / cap, which are grown by magnetron sputtering on $SiO_2$ layer on a silicon substrate [see Fig. 1(a)]. The thickness of each layer is indicated in (…) in nanometers. The thickness of the MgO layer is 1.05 nm, much thinner than those in the MTJs in our previous report (1.5 nm)[18]. The multilayer stacks are patterned into 70×200 nm² junctions by milling, then annealed in 1 T for 120 min at 330 °C to crystallize CoFeB layers.[20,21] Figure 1(b) shows a typical magnetoresistance (MR) curve at 3 K. The MTJ resistances in the P and AP configurations ($R_P$ and $R_{AP}$) are 215 Ω and 666 Ω, respectively, where the MR ratio defined by ($R_{AP}$-$R_P$)/ $R_P$ is 210 %. The area resistance (RA) is 3.01 Ω·μm². Four different MTJ devices (#1, #2, #3, and #4) were measured and all the devices gave a quantitatively consistent result.

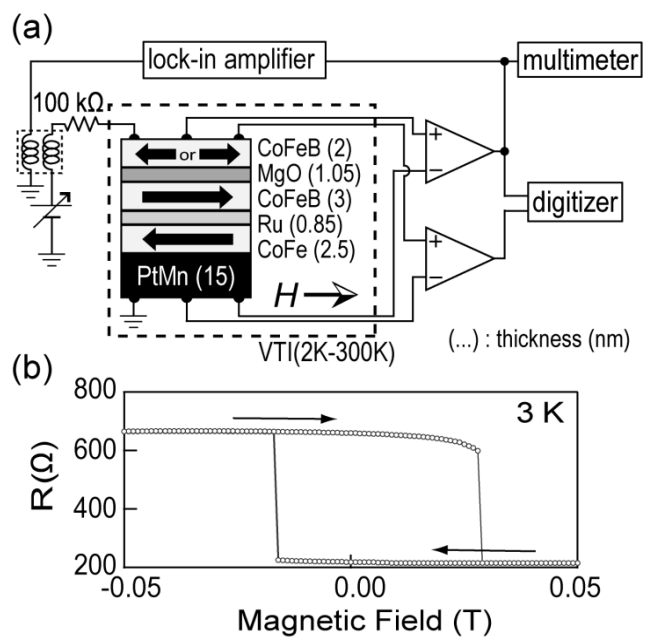

FIG. 1. (a) Present MTJs consist of buffer / PtMn(15) /CoFe(2.5) / Ru(0.85) / CoFeB(3) / MgO(1.05) / CoFeB(2) / cap multilayer. Measurement setup for the (differential) resistance and the shot noise is schematically shown. The arrow shows the direction of the external magnetic field. (b) Typical MR curves of the present MTJ measured at 3 K. The arrows show the direction of the field sweep.

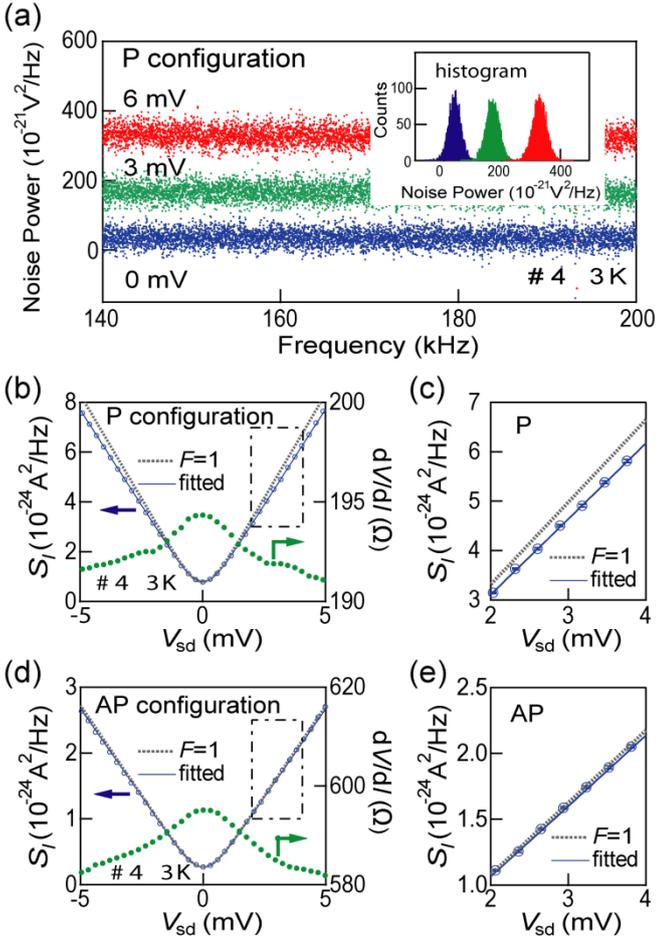

FIG. 2. (a) Measured voltage noise power spectral density of the MTJ device (#4) in the P configurations for the bias voltage ($V_{sd}$) = 0, 3, and 6 mV at 3 K. Inset shows the histograms of each spectrum between 140 and 180 kHz. (b) Differential resistance (solid mark) and estimated current noise power spectral density (open mark) for the P configuration. The solid line is the fitted curve, while the dashed line shows the curve corresponding to $F=1$. (c) A part of the graph of Fig. 2(b) (indicated by a dot-dashed rectangle) is enlarged to show that the experimental result is clearly deviated from the $F=1$ case. (d) and (e) Counterpart of Figs. 2 (b) and (c) for the AP configuration, respectively.

3, and 6 mV at 3 K in Fig. 2(a). There exists a slight resistor-capacitor (*RC*) damping due to the capacitance (760 pF) of the measurement lines. We performed the histogram analysis[18] for the data between 140 and 180 kHz (6,000 points) after taking the *RC* damping into account, which enables us to determine the noise power spectral density with the accuracy of 0.1 % [see inset in Fig. 2(a)].

The current noise power spectral density $S_I$ is obtained by $S_V=(dV/dI)^2 S_I$, where $dV/dI$ is the measured differential resistance. Figures 2(b) and (d) represent $S_I$ and $dV/dI$ for the same MTJ for the P and AP configurations at 3 K, respectively. The parabolic behavior at finite bias ($|eV_{sd}|\sim k_B T$) indicates the crossover from the thermal to shot noise. At large bias voltages ($|eV_{sd}|>>k_B T$), $S_I$ is proportional to $I$. To estimate the Fano factor, the $S_I$ is fitted to the following equations taking $V_{sd}$-dependent $dV/dI$ into account,

$$S_I = 4k_B T / \frac{dV}{dI} + 2F\left[eI \coth\left(\frac{eV_{sd}}{2k_B T}\right) - 2k_B T / \frac{dV}{dI}\right].$$

The result of the fitting is shown in solid curves with dashed curve that represent Poissonian ($F=1$) case in Figs.2 (b)-(e). Clearly, for the P configuration shown in Fig. 2(c), the shot noise is reduced from the Poissonian limit. Table I summarizes the Fano factor and MR ratio for all the four samples. The Fano factor is suppressed (F=0.91±0.01) for the P configurations, whereas it is very close to 1 (F=0.98±0.01) for the AP configurations [see Fig.2(e)]. The finite quantitative difference of the Fano factor between the two configurations, especially its reduction from unity in the P configuration, is the central result of the present work. We have confirmed that, although the non-linearity correction ($dV/dI$) is taken into account in the above analysis, the Fano factors themselves do not change when we use the resistance at $V_{sd} = 0$ mV instead of $dV/dI$ at finite $V_{sd}$ in the analysis.

TABLE I. List of Fano factors for P and AP configuration with the MR ratios and RAs at 3 K for all the measured MTJs.

| sample Nos. | Fano factor | | MR ratio (%) | RA (Ω·μm²) |
|---|---|---|---|---|
| | Parallel | Anti-parallel | | |
| 1 | 0.91±0.01 | 0.97±0.01 | 202.0 | 3.2 |
| 2 | 0.89±0.01 | 0.97±0.01 | 191.3 | 3.2 |
| 3 | 0.93±0.01 | 0.99±0.01 | 210.0 | 3.0 |
| 4 | 0.92±0.01 | 0.98±0.01 | 206.2 | 2.7 |

The MTJs were measured in the variable temperature insert (Oxford VTI) in the magnetic field (*H*) between +0.05 T and -8 T, whose direction is shown in Fig. 1(a). As schematically shown in Fig. 1(a), the DC current is applied to the MTJ through the 100 kΩ resistor. The two voltage signals across the MTJs are amplified independently by two amplifiers (NF corporation LI-75A with the input referred noise of 2 nV/$\sqrt{Hz}$ and the input impedance of 100 MΩ) at room temperature, and are recorded at a two-channel digitizer (National Instruments PCI-5922). In order to reduce the external noise, the measured two sets of time domain data are cross-correlated to yield the noise power spectral density through the fast Fourier transformation. The noise measurement system is carefully calibrated with the thermal noise by using several commercial resistors with high precision of 0.01 % (MCY100R00T, MCY250R00T, MCY350R00T, and MCY1K0000T). We have achieved the experimental precision of the Fano factor below 1% in the present setup.

The typical result of the measured voltage noise power spectral density between 140 kHz and 200 kHz is shown for the P configurations for the bias voltage ($V_{sd}$)=0,

Figure 3(a) shows that there is no dependence of the Fano factor on the magnetic field (-2~-8 T) for the P configurations at 3 K. The Fano factor is also independent of the temperature [see Fig. 3(b)], although the experimental accuracy decreases due to the 1/*f* noise contribution above 6 K. We reported that *F* is close to 1 in MTJs before[18]. For comparison, we measure the same MTJs in the present setup. As shown in Fig. 3(c), the result (*F* = 1) is reproduced within 1 %. Thus the sub-Poissonian tunneling process surely occurs in the P configuration of the present MTJ.

Ideally the electron transport process in nonmagnetic tunneling junctions (NTJs) is Poissonian with *F*=1. However, it was reported that, when there are leakage currents through localized states within the barrier, the sub-Poissonian noise can occur.[22] Regarding MTJs, Jiang *et al*. reported an observation of the Poissonian shot noise (i.e., *F*~1) in $Al_2O_3$-based MTJ's.[15] Later they reported a strong suppression of *F* (*F*~0.45).[16] The first systematic study on the shot noise was reported by Guerrero et al.,[17] who observed the Fano factor of $F\sim0.65$ in the P configuration in the amorphous $Al_2O_3$-based MTJs, and inferred a possible role of localized states in the barrier. After the above leakage current model, they assumed that nonmagnetic and/or paramagnetic impurities within the barrier form localized states, and qualitatively explained the observed spin-dependent suppression of the Fano factor. In this situation, the Fano factor is considerably smaller than unity for both P and AP configurations, reflecting the fact that the leakage via the localized states exists in both configurations.

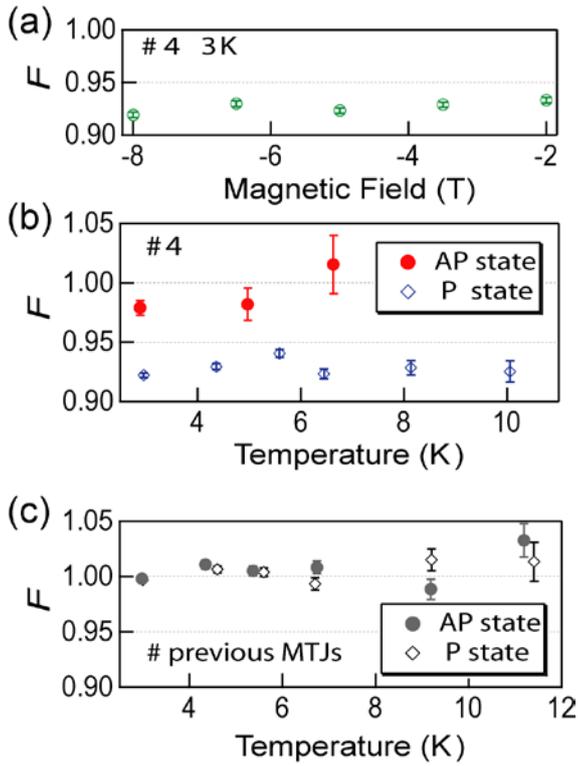

FIG. 3. (a) Dependence of the obtained Fano factor for the P configuration on the magnetic field at 3 K. The field direction is shown in Fig. 1 (a). (b) Obtained Fano factor for both P and AP configurations as a function of the sample temperature. The open and solid marks show the factors for the P and AP configurations, respectively. (c) Fano factor on the MTJ of our previous report[21] for the P and AP configurations as a function of the sample temperature.

Unlike $Al_2O_3$-based MTJs, MgO-based MTJs consist of a crystallized MgO barrier and their interface is presumed to be free from such localized states. Actually, we observed $F\sim1$ in the previous sample with the barrier thickness of 1.5 nm.[18] For the present MTJs with 1.05-nm-thick barrier, the high TMR ratio (more than 200 % at 3 K) indicates that the MTJs are well-crystallized. In addition, the Fano factor is observed to be close to 1 for the AP configuration, which can be naturally understood that the transport process is Poissonian as just expected. These facts strongly indicate that the present sub-Poissonian noise for the P configuration cannot simply be explained by the above leakage current model and is intrinsic to the electron tunneling across MTJs with a thin tunneling barrier. As the result is independent of both the magnetic field and the temperature as shown in Figs. 3(a) and (b), the possibilities of subtle effects due to magnetic impurities in the barriers and 1/f-noise contribution are safely ruled out. The theoretical prediction for the ideal interface[7] can almost reproduce the measured conductance for the P configuration. Thus, the observed sub-Poissonian process in the P configuration is intrinsic, which indicates that each tunneling event through the MTJ is anti-correlated with each other, namely "anti-bunching" of electrons occurs. We propose that the coherent transport plays a central role in this observation; In MgO-based MTJs, the coherent tunneling for the P configuration is dominated by that through a $\Delta_1$ state, whereas the $\Delta_1$ state hardly contributes to the tunneling for AP configuration.[7] The Fano factor reduction presumably reflects the difference of such tunneling processes.

In conclusion, the shot noise is measured in well-crystalline CoFeB/MgO/CoFeB-based MTJs. The obtained Fano factor for P configuration is reduced from unity (typically 0.91) indicating the sub-Poissonian process of the electron tunneling, while for the Fano factor in the AP configuration is close to 1 (typically 0.99). As the signal-to-noise (SN) ratio of MTJ devices is in principle limited by the shot noise, further understanding of the mechanism of the present sub-Poissonian tunneling would serve for constructing efficient MgO-based MTJ devices with higher SN ratio.

We appreciate fruitful comments and supports from Markus Büttiker, Wulf Wulfhekel, Eva Hirtenlechner, and Takeo Kato. This work is partially supported by JSPS Funding Program for Next Generation World-Leading Researchers.